\documentclass[prc,floatfix,groupedaddress,nofootinbib,showpacs,amsmath,amssymb,amsfonts,superscriptaddress,widetable]{revtex4-1}
\usepackage{graphicx}
\usepackage{threeparttable}
\usepackage[T1]{fontenc}

\begin{document}

\title{High-$K$ multi-particle bands and pairing reduction in $^{254}$No}

\author{Xiao-Tao He}%
\email{Corresponding author: hext@nuaa.edu.cn}
 \affiliation{College of Material Science and Technology, Nanjing University of
              Aeronautics and Astronautics, Nanjing 210016, China}
              
\author{Shu-Yong Zhao}%
 \affiliation{College of Material Science and Technology, Nanjing University of
              Aeronautics and Astronautics, Nanjing 210016, China}
              
\author{Zhen-Hua Zhang}%
 \affiliation{Mathematics and Physics Department, North China Electric Power University, Beijing 102206, China}
 
\author{Zhong-Zhou Ren}%
 \affiliation{School of Physics Science and Engineering, Tongji University, Shanghai 200092, China}

\date{\today}

\begin{abstract}
The multi-particle states and rotational properties of two-particle bands in $^{254}$No are investigated by the cranked shell model (CSM) with pairing correlations treated by a particle-number conserving (PNC) method. For the first time, the rotational bands on top of two-particle $K^{\pi}=3^+,8^-$ and $10^+$ states and the pairing reduction are studied theoretically in $^{254}$No. The experimental excitation energies and moments of inertia for the multi-particle state are reproduced well by the calculation. Better agreement with the data are achieved by including the high-order deformation $\varepsilon_{6}$ which leads to enlarged $Z=100$ and $N=152$ deformed shell gaps. The rise of the $J^{(1)}$ in these two-particle bands compared with the ground-state band is attributed to the pairing reduction due to the Pauli blocking effects.  
\end{abstract}


\keywords{Rotational band, Spin assignment, Band-crossing, High$-j$ orbital, Transfermium nuclei, Nuclear deformation}

\maketitle

\section{Introduction}

In recent years, many decay and in-beam spectroscopic studies have been performed on the light superheavy nuclei around the $Z=100, A=250$ mass region. Valuable experimental data are available to reveal the detailed structure information and to constrain various nuclear theories (see Refs.~\cite{HerzbergR2008_PPNP61,HerzbergR2004_JoPGNaPP30} and references therein). $^{254}$No is the pioneer nucleus of the experimental spectroscopy study in this mass region due to its relatively high production rate. Pioneering research includes both the extension of the ground state bands (GSB) to the high angular momentum~\cite{ReiterP1999_PRL82,ReiterP2000_PRL84,LeinoM1999_TEPJA-HaN6} and the observation of the high-$K$ multi-particle states~\cite{GhiorsoA1973_PRC7,EeckhaudtS2005_EPJA26,TandelS2006_PRL97,HerzbergR2006_N442,HesbergerF2010_EPJA43a,ClarkR2010_PLB690}. 

In 1973, a $0.28\pm0.04$ s isomer in $^{254}$No was reported, which was suspected as a $K^{\pi}=8^-$ state arising from either two-proton $\pi\frac{7}{2}^{-}[514]\otimes\pi\frac{9}{2}^+[624]$ or two-neutron $\nu\frac{9}{2}^−[734]\otimes\nu\frac{7}{2}^+[613]$ configurations~\cite{GhiorsoA1973_PRC7}. More than thirty years later, the $8^-$ isomer was identified with excitation energy from $1.293-1.297$ MeV in several experiments~\cite{TandelS2006_PRL97,HerzbergR2006_N442,HesbergerF2010_EPJA43a,ClarkR2010_PLB690}. The configuration of this state keeps still an open issue. A two-neutron state is favored in Ref.~\cite{ClarkR2010_PLB690} while a two-proton state with configuration $\pi\frac{7}{2}^{-}[514]\otimes\pi\frac{9}{2}^+[624]$ is favored in the other works~\cite{TandelS2006_PRL97,HerzbergR2006_N442,HesbergerF2010_EPJA43a}. Rotational structure on top of the $8^-$ isomer is reported independently by two contemporaneous studies, in which the detailed level schemes are proposed differently~\cite{HesbergerF2010_EPJA43a,ClarkR2010_PLB690}. He{\ss}berger \textit{et al.} suggested that all the seven observed transitions constitute a single $I = 1$ rotational sequence based on the $K^{\pi}=8^-$ state~\cite{HesbergerF2010_EPJA43a} while Clark \textit{et al.} placed only the first two members in the $K^{\pi}=8^-$ band and the rest of the transitions in the $K^{\pi}=10^+$ band. Meanwhile, the $K^{\pi}=10^+$ state with two-neutron configuration is proposed~\cite{ClarkR2010_PLB690}.    

The second isomer discovered in $^{254}$No is a four-particle state with energy $E>2.5$ MeV and half-live around~$171-198\ \mu$s~\cite{TandelS2006_PRL97,HerzbergR2006_N442,HesbergerF2010_EPJA43a,ClarkR2010_PLB690}. Its configuration can not be determined yet. $K^{\pi}=16^+$ was assumed in Refs.~\cite{HerzbergR2006_N442,HesbergerF2010_EPJA43a,ClarkR2010_PLB690} while $K^{\pi}=14^+$ was tentatively suggested by Tandel \textit{et al.}~\cite{TandelS2006_PRL97}. Note that this is one of the only two four-particle isomers reported experimentally in this region. The other one is the recently observed $247(73)\ \mu$s $K^{\pi}=16^+$ isomer in $^{254}$Rf~\cite{DavidH2015_PRL115}.

The two-particle $K^{\pi}=3^+$ state is assigned unambiguously as a two-proton state with configuration $\pi\frac{7}{2}^-[514]\otimes\pi\frac{1}{2}^-[521]$~\cite{TandelS2006_PRL97,HerzbergR2006_N442,HesbergerF2010_EPJA43a,ClarkR2010_PLB690}. The $3^{+}$ state is of particular interest since the proton $\pi\frac{1}{2}^-[521]$ orbital stems from the spherical $2f_{5/2}$ orbital. The spin-orbit interaction strength of $2f_{5/2}-2f_{7/2}$ partner governs the size of the $Z = 114$ spherical shell gap, which is predicted as the possible next magic proton number beyond lead. The properties of single-particle orbitals $\pi\frac{7}{2}^-[514]$ and $\pi\frac{1}{2}^-[521]$ effect strongly the properties of the neighboring odd-$Z$ nuclei~\cite{ChatillonA2007_PRL98,KetelhutS2009_PRL102,HeX2009_NPA817,LiY2016_SCPMA59}.

These observed high-$K$ multi-particle states in $^{254}$No can provide valuable information on the single-particle structure, deformation, pairing correlations, $K$ conservation, etc. The rotational bands built upon these multi-particle states will provide insight into the angular momentum alignment, high-$j$ intruder orbital, pairing reduction and so on. In addition, the knowledge of the transfermium nuclei can provide indirect information about the single-particle structure of the superheavy nuclei, which is crucial to the superheavy element synthesis.  

\begin{figure}[h] 
\centering
    \includegraphics[width=9cm]{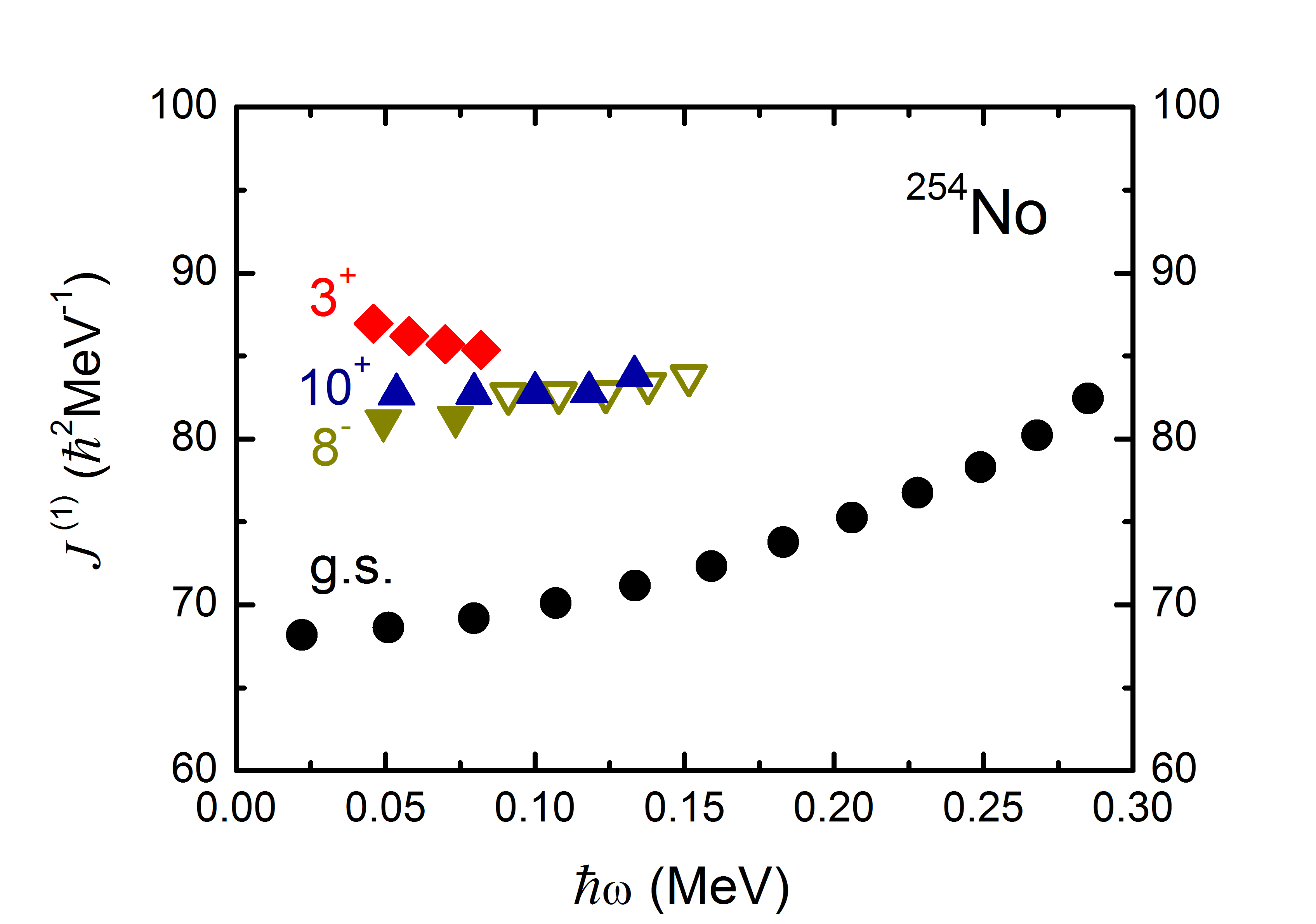}
      \caption{Experimental kinematic moments of inertia $J^{(1)}$ for the ground-state and two-particle $K^{\pi}=3^+,8^-,10^+$ state bands in $^{254}$No. The experimental data are taken from Refs.~\cite{EeckhaudtS2005_EPJA26,HerzbergR2006_N442,HesbergerF2010_EPJA43a,ClarkR2010_PLB690,nndc}. The last five transitions in the $K^{\pi}=8^-$ band in Ref.~\cite{HesbergerF2010_EPJA43a}, which are placed in the $K^{\pi}=10^+$ band in Ref.~\cite{ClarkR2010_PLB690}, are denoted by open down-triangles.}
\label{fig:Fig1}
\end{figure}

The comparison of the experimental kinematic moment of inertia (MoI) $J^{(1)}$ versus rotational frequency $\hbar\omega$ for the two-particle high-$K$ bands with the ground-state band of $^{254}$No is displayed in Fig.~\ref{fig:Fig1}. Compared with the ground-state band, a $20\%\sim25\%$ increase in $J^{(1)}$ is seen for the high-$K$ bands at the low frequency region. As the rotational frequency increasing, the ground-state band increases smoothly while the high-$K$ bands keep almost constant ($8^{-}$ and $10^{+}$ bands) or decrease ($3^{+}$ band). These behaviors will be explained by the detailed investigations about the pairing correlation,  angular momentum alignment and Pauli blocking effect, etc. 

In terms of the theoretical investigations, most of the spectroscopic studies of $^{254}$No focused on the properties of the yrast band~\cite{RenZ2002_PRC65, ZhangZ2013_PRC87, LiuH2012_PRC86, ShiY2014_PRC89, ZhangZ2012_PRC85, HeX2009_NPA817, MengX2020_SCMA63, Al-KhudairF2009_PRC79, HeX2008_IJoMPE17, ShneidmanT2006_PRC74, RenZ2003_PRC67, AfanasjevA2003_PRC67, LaftchievH2001_EPJA12, EgidoJ2000_PRL85, DuguetT2001_NPA679}. The strength of pairing correlations in $A=250$ region and its influence on the moment of inertia of the ground-state bands were compared with the lighter nuclear system in Refs.~\cite{DuguetT2001_NPA679,AfanasjevA2003_PRC67}. As for the observed high-$K$ multi-particle states, few theoretical studies have been carried out. Liu \textit{et al.} calculated the observed high-$K$ isomers in $^{254}$No with special attention paid to the influence of the high-order deformation $\beta_{6}$ on the excitation energies and the nuclear potential energy~\cite{LiuH2011_PRC83}. Jolos \textit{et al.} studied the low-lying and collective states in $Z\sim100$ nuclei with particular discussions on the effects of octupole and hexadecupole residual forces~\cite{JolosR2011_JPGNPP38}. To our best knowledge, there is still no detailed theoretical investigation of the two-particle $K^{\pi}=3^+,8^-$ and $10^+$ bands in $^{254}$No up to now.

In the present work, the multi-particle states in $^{254}$No and the rotational bands on top of them are investigated by the cranked shell model with pairing correlations treated by a particle-number conserving method. This is the first time that the detailed theoretical calculations and interpretations are performed on the observed rotational bands beyond the yrast band in $^{254}$No. Pairing correlation and blocking effects are very important to describe the multi-particle states. In the PNC-CSM method, the cranked shell model Hamiltonian with monopole and quadrupole pairing correlations is solved directly in a truncated Fock space. So the particle-number is conserved and the Pauli blocking effects are taken into account exactly. 
 
\section{Theoretical Framework}\label{sec:2}

The cranked shell model (CSM) Hamiltonian in the rotating frame is,
\begin{eqnarray}
 H_\mathrm{CSM}= H_{\rm SP}-\omega J_x + H_\mathrm{P}(0)+H_\mathrm{P}(2).
\end{eqnarray}
 $H_{\rm SP}=\sum_{\xi}(h_\mathrm{Nil})_{\xi}$ is the single-particle part, where $h_{\mathrm{Nil}}$ is the Nilsson Hamiltonian, $\xi$ ($\eta$) the eigen state of the Hamiltonian $h_{\xi(\eta)}$ and $\bar{\xi}$ ($\bar{\eta}$) the time-reversed state. $-\omega J_x$ is the Coriolis interaction with the rotational frequency $\omega$ about the $x$ axis (perpendicular to the nuclear symmetry $z$ axis). The cranked Nilsson levels $\epsilon_{\mu}$ and cranked state $|\mu\rangle$ are obtained by diagonalizing the cranked single-particle Hamiltonian $h_{0}(\omega)=h_{\xi}-\omega j_{x}$.
  
The pairing includes monopole and quadrupole pairing correlations,  
\begin{eqnarray}
 \ H_{\text{P}}(0)
 &=&-G_{0}\sum_{\xi \eta }
 a_{\xi }^{\dagger}a_{\overline{\xi }}^{\dagger }a_{\overline{\eta }}a_{\eta }\ ,
 \\
 H_{\text{P}}(2)
 &=&-G_{2}\sum_{\xi \eta } q_{2}(\xi) q_{2}(\eta)
 a_{\xi}^{\dagger } a_{\overline{\xi}}^{\dagger}
 a_{\overline{\eta}}a_{\eta}\ ,
 \label{eq:Hp}
\end{eqnarray}
where $a_{\xi }^{\dagger}a_{\overline{\xi }}^{\dagger }$ $(a_{\bar{\eta}} a_\eta)$ is the pair creation (annihilation) operator. $q_{2}(\xi) = \sqrt{{16\pi}/{5}}\langle \xi |r^{2}Y_{20} | \xi \rangle$ is the diagonal element of the stretched quadrupole operator. 

In the rotating frame, the symmetry of the time reversal is broken while the symmetry of rotation by $\pi$ around the $x$ axis, $R_{x}(\pi)=e^{-i\pi\alpha}$, is retained. The signature $\alpha=\pm1/2$,  eigenvalues of $R_{x}(\pi)$, remains a good quantum number. Transform the Hamiltonian into the cranked basis, we have,
\begin{eqnarray}
 H_\mathrm{CSM}= \sum_{\mu}\epsilon_{\mu}b_{\mu}^{\dagger}b_{\mu} 
 -G_0\sum_{\mu\mu{\prime}\nu\nu{\prime}}f^{\ast}_{\mu\mu{\prime}}f_{\nu{\prime}\nu}b_{\mu+}^{\dagger}b_{\mu{\prime}-}^{\dagger}b_{\nu-}b_{\nu{\prime}-}
-G_2\sum_{\mu\mu{\prime}\nu\nu{\prime}}g^{\ast}_{\mu\mu{\prime}}g_{\nu{\prime}\nu}
b_{\mu+}^{\dagger}b_{\mu{\prime}-}^{\dagger}b_{\nu-}b_{\nu{\prime}+},
\label{Eq:H_cranked}
\end{eqnarray}
where $b_{\mu}^{\dagger}$ is the \textit{real} particle creation operator of the cranked state $|\mu\rangle$. To investigate the pairing reduction due to rotation and blocking, particle-number conserving method (see Refs.~\cite{ZengJ1983_NPA405,WuC1989_PRC39,ZengJ1994_PRC50,XinX2000_PRC62,HeX2018_PRC98} for details) is employed to deal with the pairing correlations. The cranked shell model Hamiltonian Eq.~\ref{Eq:H_cranked} is diagonalized in a truncated Cranked Many-Particle Configuration (CMPC) space~\cite{WuC1989_PRC39}. The effective pairing strengths $G_0$ and $G_2$ are connected with the dimension of the truncated CMPC space. In the following calculations, the CMPC space for $^{254}$No is constructed in the proton $N= 4, 5, 6$ and neutron $N= 6, 7$ shells. The dimensions of the CMPC space are about 1000 and the corresponding effective monopole and quadrupole pairing strengths are $G_{0} = 0.25$ MeV and $G_{2} = 0.02$ MeV for both protons and neutrons, which are determined by the odd-even differences in moment of inertia in this mass region. Since the total Hamiltonian is diagonalized directly in a truncated Fock space, the sufficiently accurate solutions can be obtained in a comparatively small diagonalization space for the yrast and low-lying excited states. By this way, like the standard shell-model approach, the particle-number keeps conserved and the Pauli blocking effect is taken into account exactly.   


The eigenstate of $H_{\textrm{CSM}}$ is $| \psi \rangle = \sum_{i} C_i | i \rangle$ with CMPC $| i \rangle$ defined by the occupation of real particles on the cranked single-particle orbitals. The converged solution $| \psi \rangle$ can always be obtained even for a pair-broken state while the conventional cranked Hartree-Fock-Bogoliubov model does not in many cases~\cite{FuX2013_PRC87,FuX2014_PRC89}. This makes it very convenient to treat the multi-particle states in a nucleus~\cite{HeX2018_PRC98}. The PNC-CSM method provides a reliable way to assign the configuration for a multi-particle state. Once the wave function $| \psi \rangle$ is obtained, the configurations for all the low-lying excitation multi-particle states can be obtained by the occupation probability of the specific $| i \rangle$ with unpaired particle blocked in the single-particle orbitals near the Fermi surface. 

\section{Nilsson single-particle levels}\label{sec:Nilsson}

\begin{figure}[h] 
\centering
    \includegraphics[width=12cm]{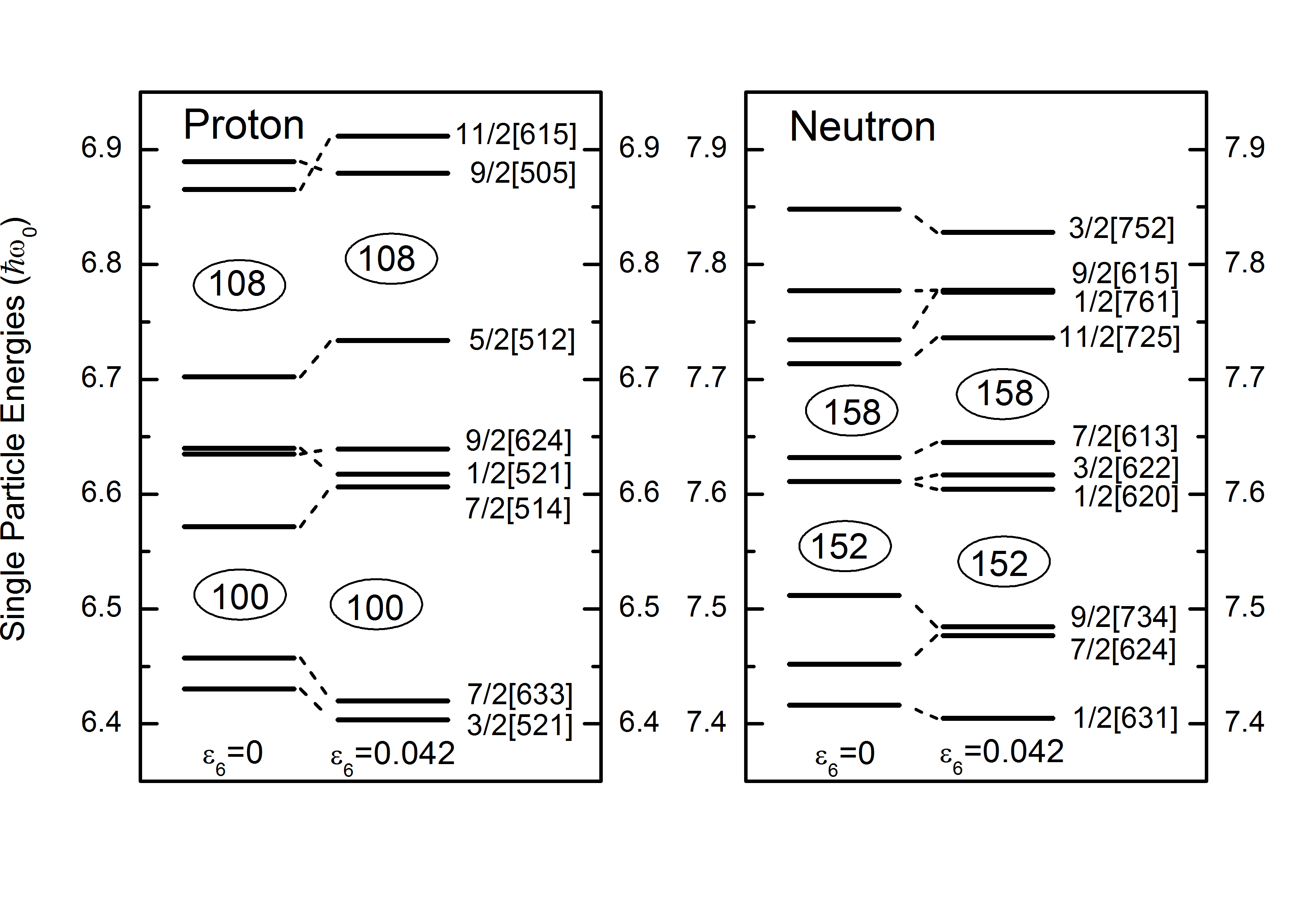}
      \caption{Nilsson levels near the Fermi surface of $^{254}$No. The deformation parameters are $\varepsilon_{2}=0.26$, $\varepsilon_{4}=0.02$, $\varepsilon_{6}=0.0$ (left column) and $\varepsilon_{6}=0.042$ (right column).}
\label{fig:Fig2}
\end{figure}

The Nilsson parameters $(\kappa,\mu)$, which were optimized to reproduce the experimental level schemes for light superheavy nuclei around $A=250$ mass region in Refs.~\cite{ZhangZ2011_PRC83,ZhangZ2012_PRC85}, are used in this work. The values of proton $\kappa_{5},\mu_{5}$ and neutron $\kappa_{6},\mu_{6}$ are modified slightly to reproduce the correct single-particle level sequence when $\varepsilon_{6}$ is included. The deformation parameters $\varepsilon_{2}=0.26,\ \varepsilon_{4}=0.02$ are taken from Ref.~\cite{ZhangZ2012_PRC85} and $\varepsilon_{6}=0.042$ is taken from Ref.~\cite{MoellerP1995_ADaNDT59}. 

The Nilsson single-particle levels with and without high-order deformation $\varepsilon_{6}$ are compared at rotational frequency $\hbar\omega=0$ in Fig.~\ref{fig:Fig2}. It can be see that the calculation including $\varepsilon_{6}$ deformation leads to enlarged proton $Z=100$ and neutron $N=152$ deformed shell gaps, which is consistent with predictions of Woods-Saxon potential calculations by Liu \textit{et al.}~\cite{LiuH2011_PRC83} and Patyk \textit{et al.}~\cite{PatykZ1991_NPA533}. Note that the existence of these two deformed shell gaps have been confirmed by the experiment~\cite{GreenleesP2008_PRC78}. In addition, compared with the results without $\varepsilon_{6}$ deformation, the deformed shell gap at proton $Z=106$ becomes larger and the one at $Z=108$ becomes smaller, shell gaps at neutron $N=148,160$ appear and the one at $N=150$ disappears. The changes of the deformed single-particle level structure will further influence the excitation energy and the moment of inertia of the multi-particle states. 

The effect of the $\varepsilon_{6}$ deformation on the multi-particle states in $^{254}$No has been investigated in detail by the configuration-constrained potential-energy surfaces (PES) calculations~\cite{LiuH2011_PRC83}. The authors stated that by including the $\varepsilon_{6}$ deformation, the multi-particle states gain extra binding energies. Therefore they will have an enhanced stability against fission. This conclusion is confirmed by the present PNC-CSM calculations. However, the influence of the high-order deformation is still intricate, especially in heavy and superheavy mass region where the single-particle level density is high and the knowledge of single-particle level structure is limited. Moreover, the value of $\varepsilon_{6}$ is strongly model dependent. Therefore, more comprehensive investigation into the $\varepsilon_{6}$ deformation effect on the single-particle levels is needed in heavy and superheavy nuclei mass region.

\section {\label{subsec:states}Multi-particle states} 

\begin{figure}[h] 
\centering
    \includegraphics[width=14cm]{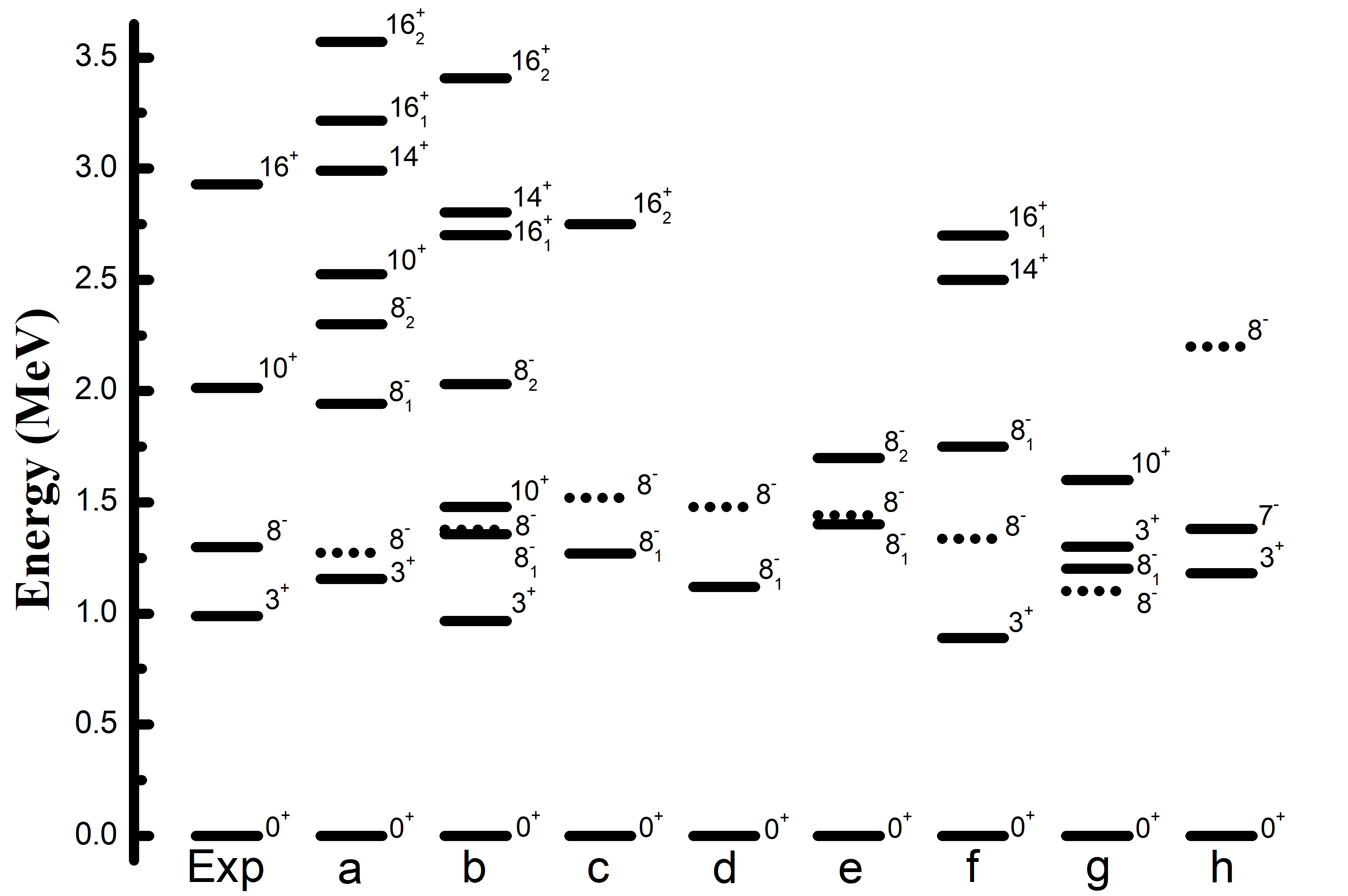}
      \caption{Comparison between the excitation energies of the experimentally deduced and calculated multi-particle states in $^{254}$No. 
     The calculations come from 
     a) PNC-CSM method in this work; 
     b) configuration-constrained potential-energy surfaces method~\cite{LiuH2011_PRC83}; 
     c) projected shell model~\cite{HerzbergR2006_N442}; 
     d) configuration-constrained potential-energy surfaces method~\cite{XuF2004_PRL92}; 
     e) quasiparticle-phonon nuclear model~\cite{SolovievV1991_SJNP54}; 
     f) Woods-Saxon potential plus a Lipkin-Nogami formalism for pairing~\cite{TandelS2006_PRL97}; 
     g) semi-microscopic approach~\cite{IvanovaS1976_SJPN7}; 
     h) Skyrme Hartree-Fock Bogolyubov model with the SLy4 force~\cite{TandelS2006_PRL97}. 
     This plot extends a similar figure shown in Refs.~\cite{HerzbergR2006_N442,ClarkR2010_PLB690}. 
     The $8^-$ state with two-proton configuration $\pi\frac{7}{2}^{-}[514]\otimes\pi\frac{9}{2}^{+}[624]$ is indicated by dot lines. 
     The $8^-_1$ stands for the two-neutron state with configuration $\nu\frac{9}{2}^{-}[734]\otimes\nu\frac{7}{2}^{+}[613]$ and 
     $8^-_2$ for $\nu\frac{9}{2}^{-}[734]\otimes\nu\frac{7}{2}^{+}[624]$. 
     The $16^+_1$ stands for the four-particle state with configuration $\nu\frac{9}{2}^{-}[734]\otimes \nu\frac{7}{2}^{+}[613]\otimes \pi\frac{7}{2}^{-}[514]\otimes\pi\frac{9}{2}^{+}[624]$ and 
     $16^+_2$ for $\nu\frac{9}{2}^{-}[734]\otimes \nu\frac{7}{2}^{+}[624]\otimes\pi\frac{7}{2}^{-}[514]\otimes\pi\frac{9}{2}^{+}[624]$.}
\label{fig:Fig3}
\end{figure}

\begin{table*}
\begin{threeparttable}
\centering
\caption{Low-lying multi-particle states in $^{254}$No predicted by the PNC-CSM calculations.}
\begin{ruledtabular}
\begin{tabular}{ccccc}
 $K^\pi$ & Configuration  & $E_x$(MeV)($\varepsilon_{6}\neq0$) &$E_x$(MeV)($\varepsilon_{6}=0$)& $E_{x}^{exp}$(MeV) \\
 \hline
    $\underline{3}^+,4^+$   &$\pi\frac{7}{2}^{-}[514]\otimes\pi\frac{1}{2}^{-}[521]$     &1.154 & 1.508 &0.988\\
    $\underline{8}^-,1^-$   &$\pi\frac{7}{2}^{-}[514]\otimes\pi\frac{9}{2}^{+}[624]$     &1.272 & 1.431 &1.297\\
    $\underline{5}^-,4^-$   &$\pi\frac{9}{2}^{+}[624]\otimes\pi\frac{1}{2}^{-}[521]$     &1.324& 1.749 &\\
    $\underline{6}^+,1^+$   &$\pi\frac{7}{2}^{-}[514]\otimes\pi\frac{5}{2}^{-}[512]$     &1.794& 1.807 &\\
    $\underline{3}^+,2^+$   &$\pi\frac{5}{2}^{-}[512]\otimes\pi\frac{1}{2}^{-}[521]$     &1.902& 2.235 &\\
    $7^-,\underline{2}^-$   &$\pi\frac{9}{2}^{+}[624]\otimes\pi\frac{5}{2}^{-}[512]$     &2.007& 2.142 &\\
    $\underline{4}^-,3^-$   &$\pi\frac{7}{2}^{+}[633]\otimes\pi\frac{1}{2}^{-}[521]$     &2.200& 2.145 &\\
    $\underline{7}^-,0^-$   &$\pi\frac{7}{2}^{+}[633]\otimes\pi\frac{7}{2}^{-}[514]$     &2.229&    &\\
    $\underline{2}^+,1^+$   &$\pi\frac{1}{2}^{-}[521]\otimes\pi\frac{3}{2}^{-}[521]$     &2.279&    &\\
    $\underline{4}^-,5^-$   &$\nu\frac{9}{2}^{-}[734]\otimes\nu\frac{1}{2}^{+}[620]$     &1.686& 1.678 &     \\
    $\underline{6}^-,3^-$   &$\nu\frac{9}{2}^{-}[734]\otimes\nu\frac{3}{2}^{+}[622]$     &1.718& 1.675  &\\
    $\underline{4}^+,3^+$   &$\nu\frac{7}{2}^{+}[624]\otimes\nu\frac{1}{2}^{+}[620]$     &1.757& 2.142 &\\
    $5^+,\underline{2}^+$   &$\nu\frac{7}{2}^{+}[624]\otimes\nu\frac{3}{2}^{+}[622]$     &1.793& 2.145 &\\
    $8^-,\underline{1}^-$   &$\nu\frac{9}{2}^{-}[734]\otimes\nu\frac{7}{2}^{+}[613]$     &1.944& 1.848 &1.297\\
    $\underline{7}^+,0^+$   &$\nu\frac{7}{2}^{+}[624]\otimes\nu\frac{7}{2}^{+}[613]$     &2.025& 2.303 &\\
    $\underline{2}^+,1^+$   &$\nu\frac{1}{2}^{+}[620]\otimes\nu\frac{3}{2}^{+}[622]$     &2.277&    &\\
    $\underline{8}^-,1^-$   &$\nu\frac{9}{2}^{-}[734]\otimes\nu\frac{7}{2}^{+}[624]$     &2.301 & 2.286  &\\
    $\underline{2}^+,3^+$   &$\nu\frac{1}{2}^{+}[620]\otimes\nu\frac{5}{2}^{+}[622]$     &2.421& 2.517 &\\
    $\underline{1}^+,0^+$   &$\nu\frac{1}{2}^{+}[620]\otimes\nu\frac{1}{2}^{+}[631]$     &2.448&    &\\
    $\underline{3}^+,4^+$   &$\nu\frac{1}{2}^{+}[620]\otimes\nu\frac{7}{2}^{+}[613]$     &2.470&    &\\
    $\underline{4}^+,1^+$   &$\nu\frac{5}{2}^{+}[622]\otimes\nu\frac{3}{2}^{+}[622]$     &2.499&    &\\
    $10^+,\underline{1}^+$  &$\nu\frac{9}{2}^{-}[734]\otimes\nu\frac{11}{2}^{-}[725]$    &2.526 & 2.454  &2.013\\
    $14^+$  &$\nu\frac{9}{2}^{-}[734]\otimes\nu\frac{3}{2}^{+}[622]\otimes\pi\frac{7}{2}^{-}[514]\otimes\pi\frac{9}{2}^{+}[624]$     &2.991&     &2.928\\
    $16^+$   &$\nu\frac{5}{2}^{-}[523]\otimes\nu\frac{7}{2}^{+}[613]\otimes\pi\frac{7}{2}^{-}[514]\otimes\pi\frac{9}{2}^{+}[624]$   &3.215&     &2.928\\
    $16^+$   &$\nu\frac{9}{2}^{-}[734]\otimes\nu\frac{7}{2}^{+}[624]\otimes\pi\frac{7}{2}^{-}[514]\otimes\pi\frac{9}{2}^{+}[624]$   &3.572&     &\\
\end{tabular}
\end{ruledtabular}
\label{tab:states_No254}
\begin{tablenotes}
\item[1] The $K^{\pi}$ values favored by the GM rules~\cite{GallagherC1962_PR126} are underlined for each two-particle state.
\end{tablenotes}
\end{threeparttable}
\end{table*} 

The multi-particle states predicted by various models are compared with the experimental data in Fig.~\ref{fig:Fig3}. The more comprehensive predictions of PNC-CSM calculation are listed in Table~\ref{tab:states_No254}. Our model in its present version does not include the residual spin-spin interaction. In Table~\ref{tab:states_No254}, both $K^{\pi}=|\Omega_1\pm\Omega_2|^\pi$ values are shown for the two-particle states, with the value favored by the Gallagher-Moszkowski rules~\cite{GallagherC1962_PR126} underlined. According to GM rules, the spin singlet coupling is energetically favored for the pair-broken states in an even-even nucleus. 

The two-particle state at $0.988$ MeV is firmly assigned as the two-proton $3^+$ state with configuration $\pi\frac{1}{2}^{-}[521]\otimes\pi\frac{7}{2}^{-}[514]$~\cite{ClarkR2010_PLB690,TandelS2006_PRL97,HesbergerF2010_EPJA43a,HerzbergR2006_N442}. Thus this firm assignment can be used to constrain the parameterizations of theoretical models. As shown in Fig.~\ref{fig:Fig3}, the $3^+$ state is predicted as the lowest two-particle state in the present PNC-CSM calculation, configuration-constrained calculations of potential-energy surfaces~\cite{LiuH2011_PRC83}, Woods-Saxon potential plus a Lipkin-Nogami formalism for pairing~\cite{TandelS2006_PRL97} and the Skyrme Hartree-Fock Bogolyubov (SHFB) model with the SLy4 force~\cite{TandelS2006_PRL97}.  

In Table.~\ref{tab:states_No254}, the effect of the high-order deformation $\varepsilon_{6}$ on the excitation energies of the multi-particle states is demonstrated. Calculations without $\varepsilon_{6}$ lead to the result that the $8^{-}$ ($\pi\frac{9}{2}^+[624]\otimes\pi\frac{7}{2}^-[514]$), instead of the $3^+$ ($\pi\frac{1}{2}^{-}[521]\otimes\pi\frac{7}{2}^{-}[514]$) state, is the lowest two-particle state, which disagrees with the experimental result. When $\varepsilon_{6}=0.042$ is considered, the $3^{+}$ state becomes the lowest-lying two-particle state and the calculated energy reproduces the experimental data very well. This is because once $\varepsilon_{6}$ is included, proton orbitals $\pi\frac{7}{2}^{-}[514]$ and $\pi\frac{1}{2}^{-}[521]$ will get closer, and the positions of $\pi\frac{1}{2}^{-}[521]$ and $\pi\frac{9}{2}^{+}[624]$ orbitals will be reversed (see Fig.~\ref{fig:Fig2}). Besides the $3^+$ state, in general, the theoretical results with $\varepsilon_{6}$ reproduce the experimental data better for other multi-particle states as well. This is evidence that including the $\varepsilon_{6}$ leads to a more reasonable single-particle level structure for this mass region. 

The $K^{\pi}=8^{-}$ isomer is observed systematically in this mass region. Unlike the $8^{-}$ isomer in the $N=150$ isotones, its configuration is accepted as a two-neutron state with configuration $\nu\frac{9}{2}^{-}[734]\otimes\nu\frac{7}{2}^{+}[624]$ in $^{252}$No~\cite{SulignanoB2012_PRC86},  $^{250}$Fm~\cite{GreenleesP2008_PRC78} and $^{244}$Pu~\cite{HotaS2016_PRC94} in the literature, the configuration of the observed $8^{-}$ isomer at $1.297$ MeV in $^{254}$No is in dispute up to now. The two-neutron configuration is favored by the most recent experiment study~\cite{ClarkR2010_PLB690} whereas the two-proton configuration $\pi\frac{7}{2}^{-}[514]\otimes\pi\frac{9}{2}^{+}[624]$ is suggested in the earlier experimental works~\cite{TandelS2006_PRL97,HesbergerF2010_EPJA43a,HerzbergR2006_N442}. 

Theoretically, the Skyrme Hartree-Fock Bogolyubov model with the SLy4 force gives only one low-lying $8^-$ state with two-proton configuration, and it is too high in energy. All the calculations by macroscopic-microscopic (MM) method predict at least two low-lying $8^-$ states with similar excitation energies. One is the two-proton state and the other one is the two-neutron state. The calculations of the projected shell model (PSM)~\cite{HerzbergR2006_N442,HaraK1995_IJoMPE04} and the quasiparticle-phonon nuclear model~\cite{SolovievV1991_SJNP54} favor the two-neutron configuration for the lowest-lying $8^-$ state. In contrast, other MM methods, including the present PNC-CSM, Woods-Saxon plus a Lipkin-Nogami treatment for pairing~\cite{TandelS2006_PRL97}, the configuration-constrained potential-energy surfaces~\cite{LiuH2011_PRC83} and the semi-microscopic approach~\cite{IvanovaS1976_SJPN7} calculations favor the two-proton configuration assignment. The study of the configuration-constrained calculations of potential-energy surfaces leads to a lowest two-neutron $8^-$ in the earlier work~\cite{XuF2004_PRL92}. However, when the high-order $\varepsilon_{6}$ deformation is included, the proton configuration, instead of the neutron configuration, is assigned to the lowest-lying $8^-$ state~\cite{LiuH2011_PRC83}. In the present PNC-CSM calculation, three low-lying $8^-$ states are predicted. The lowest $8^{-}$ state is the two-proton state with configuration $\pi\frac{9}{2}^{+}[624]\otimes\pi\frac{7}{2}^{-}[514]$ at energy $1.272$ MeV ($\varepsilon_{6}=0.042$), which reproduces the experimental data of 1.297 MeV very well. The predicted low-lying two-neutron $8^{-}$ states are $\nu\frac{9}{2}^{-}[734]\otimes\nu\frac{7}{2}^{+}[613]$ (denoted as $8^{-}_{1}$) and $\nu\frac{9}{2}^{-}[734]\otimes\nu\frac{7}{2}^{+}[624]$ (denoted as $8^{-}_{2}$) configuration states. The latter is too high in the energy to be the observed isomer. Since the $8^{-}_{1}$ state is not the energetically favored one of the GM doublet, the excitation energy would be even higher when considering the residual spin-spin interaction. However, the $8^{-}_{1}$ state can not be completely excluded when we study its rotational behavior, which will be discussed in the next section.  

Four-particle isomer formed by coupling the two-proton and two-neutron states was observed in $^{254}$No. Two possible spin-parity assignments, i.e., $K^{\pi}=16^{+}$ and $K^{\pi}=14^{+}$, were suggested in Refs.~\cite{ClarkR2010_PLB690,HesbergerF2010_EPJA43a,HerzbergR2006_N442} and Ref.~\cite{TandelS2006_PRL97}, respectively. The present PNC-CSM calculations predict one $14^+$ state and two $16^+$ states. As shown in Fig~\ref{fig:Fig3}, $K^{\pi} = 14^{+}$ state with configuration $\nu\frac{9}{2}^{-}[734]\otimes \nu\frac{3}{2}^{+}[622]\otimes \pi\frac{7}{2}^{-}[514]\otimes\pi\frac{9}{2}^{+}[624]$ reproduces the experimental data very well. The lower $K^{\pi}=16^+_{1}$ state with the configuration of $\nu\frac{9}{2}^{-}[734]\otimes \nu\frac{7}{2}^{+}[613]\otimes \pi\frac{7}{2}^{-}[514]\otimes\pi\frac{9}{2}^{+}[624]$ is higher than the experimental data by about 0.287 MeV. The deviation is acceptable, and this configuration is favored by the most recent experimental work~\cite{ClarkR2010_PLB690} and the Wood-Saxon potential calculation~\cite{LiuH2011_PRC83}. Therefore, neither the $K^{\pi} = 16^+_1$ state or the $K^{\pi} = 14^{+}$ state can be ruled out by the present calculations. The excitation energy of the second $K^{\pi}=16^+_2$ state with configuration $\nu\frac{9}{2}^{-}[734]\otimes \nu\frac{7}{2}^{+}[624]\otimes\pi\frac{7}{2}^{-}[514]\otimes\pi\frac{9}{2}^{+}[624]$ is much larger than the experimental data, which is too high to be the observed four-particle isomer. 

It can be seen in Table~\ref{tab:states_No254} that all the three four-particle states are built on coupling different two-neutron states with the same two-proton $\pi\frac{7}{2}^{-}[514]\otimes\pi\frac{9}{2}^{+}[624]$ state. Therefore, the main uncertainty is brought from the two-neutron states. The neutron single-particle level density is very high and their structure is complicated in the heavy and superheavy mass region. Different potential will result in quite different single-particle level structure, which is very sensitive to the adopted parameters. Therefore, a further investigation into the single-particle level structure, especially for neutrons, is urgent in this mass region.  

The $K^{\pi}=10^+$ state was reported in the most recent experiment~\cite{ClarkR2010_PLB690}. As shown in Fig.~\ref{fig:Fig3}, $10^+$ state is predicted by calculations of the PNC-CSM method, configuration constrained PES~\cite{LiuH2011_PRC83} and the semi-microscopic approach~\cite{IvanovaS1976_SJPN7}. The latter two calculations are based on the Woods-Saxon single-particle levels, of which there is a deformed shell gap at neutron $N=162$ and the neutron orbital $\nu\frac{11}{2}^{-}[725]$ locates below this gap~\cite{IvanovaS1976_SJPN7,ChasmanR1977_RMP49}. In contrast, PNC-CSM calculation is based on the Nilsson single-particle levels. It differs from the Woods-Saxon potential, as shown in Fig.~\ref{fig:Fig2}, a deformed shell gap appears at neutron $N=158$, and the $\nu\frac{11}{2}^{-}[725]$ level locates just above this gap. Moreover, including of $\varepsilon_{6}$ makes the $N=158$ deformed shell gaps even larger, which results in the rise of excitation energies for both $K^{\pi}=8^{-}$ and $K^{\pi}=10^{+}$ states. Based on such single-particle level structure, the excitation energy of the $K^{\pi}=10^{+}$ state given by Nilsson potential in present calculation is 0.513 MeV higher than the experimental data whereas the results given by Woods-Saxon potential are 0.534 and 0.413 MeV lower than the experimental data in Ref.~\cite{LiuH2011_PRC83} and Ref.~\cite{IvanovaS1976_SJPN7}, respectively. It should be noted that the $K^{\pi}=10^+$ coupling is not the energetically favored one of the GM doublet. When considering the residual spin-spin interaction, the excitation energy would be higher.  

\section {\label{subsec:MOI}Moments of inertia}

\begin{figure}[h] 
\centering
    \includegraphics[width=13cm]{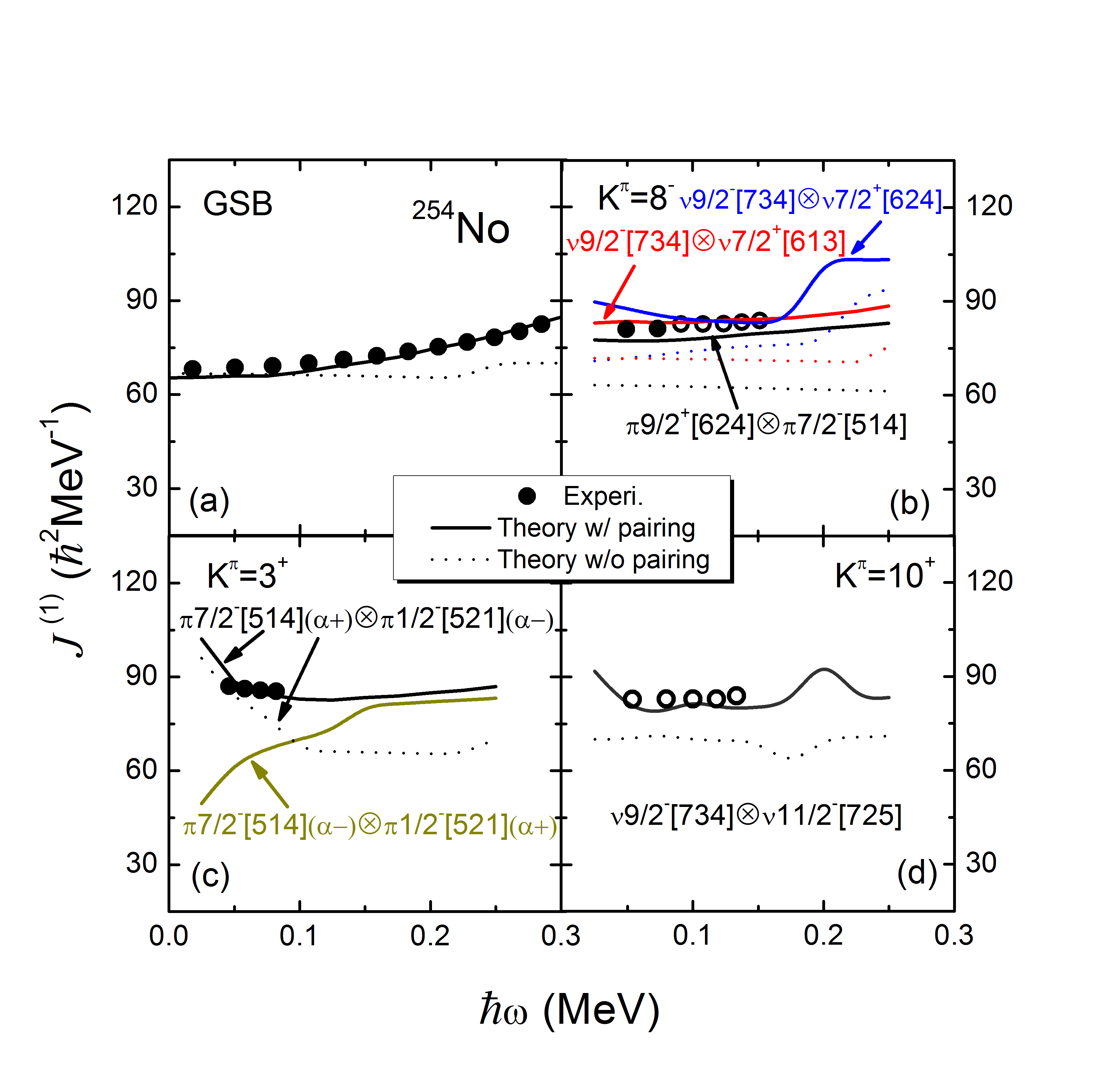}
      \caption{Kinematic moments of inertia $J^{(1)}$ versus rotational frequency for the ground-state and two-particle ($K^{\pi}=3^+,8^-$ and $10^+$) bands in $^{254}$No. Experimental data~\cite{EeckhaudtS2005_EPJA26,HerzbergR2006_N442,HesbergerF2010_EPJA43a,ClarkR2010_PLB690,nndc} are denoted by symbols and theoretical results with/without pairing are denoted by solid/dot lines. The last five transitions in the $K^{\pi}=8^-$ band~\cite{HesbergerF2010_EPJA43a}, which are placed in the $K^{\pi}=10^+$ band~\cite{ClarkR2010_PLB690}, are denoted by open circles.}
\label{fig:Fig4}
\end{figure}

The kinematic moment of inertia of the state $\left\vert \psi \right\rangle$ is given by $J^{(1)}=\left\langle \psi \right\vert J_{x}\left\vert\psi \right\rangle/\omega$, where the angular momentum alignment is $\left\langle \psi \right| J_{x}
 \left| \psi \right\rangle=\sum_{i}\left|C_{i}\right| ^{2}
   \left\langle i\right| J_{x}\left| i\right\rangle
 + 2\sum_{i<j}C_{i}^{\ast }C_{j}
   \left\langle i\right| J_{x}\left| j\right\rangle$. The calculated $J^{(1)}$ versus rotational frequency based on the ground-state and two-particle $K^{\pi}=3^+,8^-$ and $10^+$ states in $^{254}$No are compared with the experimental data in Fig.~\ref{fig:Fig4}. In general, the experimental data are reproduced quite well. 
   
The $3^{+}$ state is of particular interesting since the $\pi\frac{1}{2}^{-}[521]$ orbital originates from the spherical $2f_{5/2}$ orbital. The spin-orbit interaction strength of $2f_{5/2}-2f_{7/2}$ partner controls whether the proton $Z = 114$ is a magic number for the "island of stability" for shell stabilized superheavy nuclei. Rotational bands based on $\pi\frac{1}{2}^{-}[521]$ orbital have been observed in odd-proton nuclei $^{251}$Md~\cite{ChatillonA2007_PRL98} and $^{255}$Lr~\cite{KetelhutS2009_PRL102}. Studies of these rotational bands found a significant signature splitting~\cite{HeX2009_NPA817,LiY2016_SCPMA59}. The result of the $3^{+}$ band in $^{254}$No is similar. While the bandhead energy of the $\pi\frac{7}{2}^{-}[514](\alpha=+1/2)\otimes\pi\frac{1}{2}^{-}[521](\alpha=-1/2)$ band is only lower than the $\pi\frac{7}{2}^{-}[514](\alpha=-1/2)\otimes\pi\frac{1}{2}^{-}[521](\alpha=+1/2)$ band by about 0.6 keV, the rotational behavior is quite different. As shown in Fig.~\ref{fig:Fig4} (c), only the former can reproduce the experimental data well.

For the $8^{-}$ band, as shown in Fig.~\ref{fig:Fig4} (b), the calculated $J^{(1)}$ of the $\nu\frac{9}{2}^{-}[734]\otimes\nu\frac{7}{2}^{+}[624]$ band can not reproduce the increasing trend of the experimental data. In Ref.~\cite{ClarkR2010_PLB690}, there are only two exited members in the $K^{\pi}=8^-$ band, which are denoted by the solid circles in Fig.~\ref{fig:Fig4} (b). In this case, the experimentally deduced $J^{(1)}$ just locates between the theoretical two-neutron $\nu\frac{9}{2}^{-}[734]\otimes\nu\frac{7}{2}^{+}[613]$ and two-proton $\pi\frac{7}{2}^{-}[514]\otimes\pi\frac{9}{2}^{+}[624]$ bands. In Ref.~\cite{HesbergerF2010_EPJA43a}, the $K^\pi=8^-$ is extended to spin $I=15\hbar$. The corresponding data of the possible $K^\pi=8^-$ band extension [open circles in Fig.~\ref{fig:Fig4} (b)] are placed in the $K^\pi=10^+$ band in Ref.~\cite{ClarkR2010_PLB690} [open circles in Fig.~\ref{fig:Fig4} (d)]. If the possible $K^{\pi}=8^-$ band extension is considered, the calculated two-neutron $\nu\frac{9}{2}^{-}[734]\otimes\nu\frac{7}{2}^{+}[613]$ band agrees better with the experimental data. But overall, although the calculated $J^{(1)}$ of the $\pi\frac{7}{2}^{-}[514]\otimes\pi\frac{9}{2}^{+}[624]$ band is a bit lower than the experimental data, it is still good enough. The underestimation of the calculation may come from the influence of the effective pairing strengths. Analyzed together with the result of the excitation energies, neither the proton configuration $\pi\frac{7}{2}^{-}[514]\otimes\pi\frac{9}{2}^{+}[624]$ or the neutron configuration $\nu\frac{9}{2}^{-}[734]\otimes\nu\frac{7}{2}^{+}[613]$ can be ruled out. Further investigations are needed from both experimental and theoretical sides. 

For the $K^\pi=10^+$ band, as shown in the last section, it is not a very low excitation state in the present calculations. The occupation of the $\nu\frac{9}{2}^{-}[734]\otimes\nu\frac{11}{2}^{-}[725]$ configuration is less pure. Comparatively large probability amplitude of other components in the wave function influences the behavior of the $K^\pi=10^+$ band. Like the hump at $\hbar\omega\approx0.2$ MeV, it is attributed to the contribution from the $\nu\frac{9}{2}^{-}[734]\otimes\nu\frac{1}{2}^{-}[761]$ configuration. 

As shown in Fig.~\ref{fig:Fig1}, compared with the ground-state band, the rotational bands based on the three two-particle states $(K^{\pi}=3^+, 8^-,10^+)$ increase in $J^{(1)}$ by about $25\%$ at low frequency region. A similar rise is seen for two-particle bands in the $A=180$ region and it has been attributed to the pairing reduction~\cite{DracoulisG1998_PLB419}. To examine whether the increase of $J^{(1)}$ comes from the pairing reduction of the high-$K$ bands in $^{254}$No, $J^{(1)}$ is calculated without pairing, which is shown by dot lines in Fig.~\ref{fig:Fig4}. It shows that all the three two-particle bands have similar $J^{(1)}$ values, i.e. 65-70 $\hbar^{2}$MeV$^{-1}$, which is almost equal to the ground-state band. $J^{(1)}$ based on the ground-state keeps almost constant with frequency $\hbar\omega$ when pairing is not included. Thus, we conclude that the rise of $J^{(1)}$ for the high-$K$ bands comparing with the ground-state band at low frequency and the gradual increase in $J^{(1)}$ versus frequency of the ground-state band are mainly attributed to the pairing reduction. 
 
\section {\label{subsec:pairing}Pairing correlations}

\begin{figure}[h] 
\centering
    \includegraphics[width=11cm]{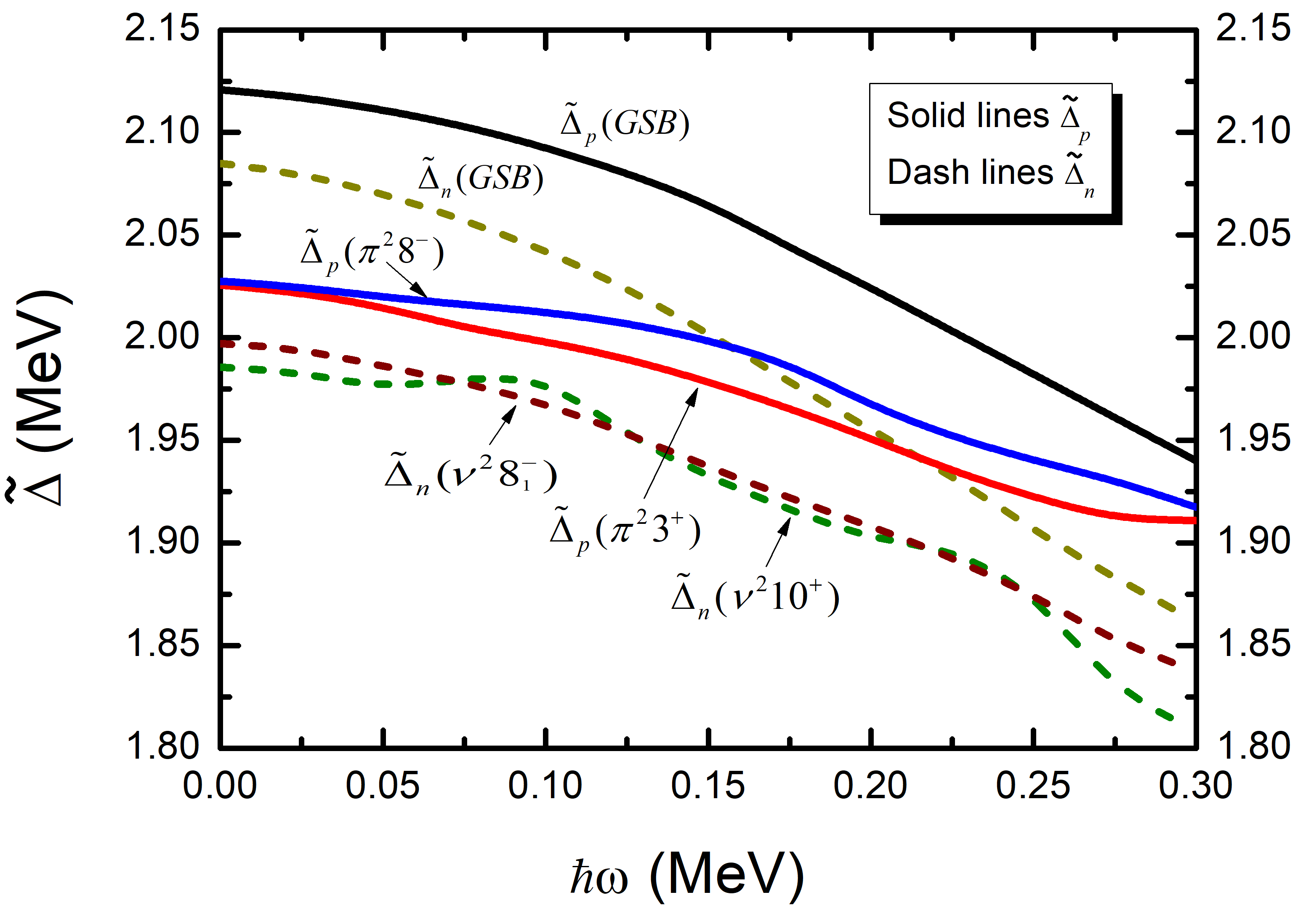}
      \caption{Calculated pairing gaps $\tilde{\Delta}$ for the ground-state and two-particle bands in $^{254}$No. The configurations of two-particle bands are $\pi^2 8^- \{\frac{9}{2}^+[624]\otimes\frac{7}{2}^-[514]\}$, $\pi^2 3^+ \{\frac{1}{2}^-[521]\otimes\frac{7}{2}^-[514]\}$, $\nu^2 8^-_1 \{\frac{9}{2}^-[734]\otimes\frac{7}{2}^+[613]\}$ and $\nu^2 10^+ \{\frac{9}{2}^-[734]\otimes\frac{11}{2}^-[725]\}$.}
\label{fig:Fig5}
\end{figure}

The nuclear pairing gap~\cite{ShimizuY1989_RMP61,WuX2011_PRC83} in the PNC-CSM formalism is defined as,
\begin{equation}
\tilde{\Delta}=G_{0}\left[ -\frac{1}{G_{0}}\left\langle\psi\right| H_{\mathrm{P}}\left|\psi\right\rangle\right]^{1/2}.
\label{eq:pairing}
\end{equation} 
For the quasi-particle vacuum band, $\tilde{\Delta}$ is reduced to the usual definition of the nuclear pairing gap $\Delta$~\cite{WuX2011_PRC83}. Figure~\ref{fig:Fig5} shows the calculated neutron and proton pairing gaps $\tilde{\Delta}$ versus rotational frequency for the ground-state band and two-particle $K^{\pi}=3^+,8^-$ and $10^+$ bands in $^{254}$No. The effective pairing strength parameters in the calculation are same for neutrons and protons. The difference in the pairing gaps between neutrons and protons comes purely from the wave functions. In general, as shown in Fig.~\ref{fig:Fig5}, the pairing gaps of neutrons are larger than that of protons. The pairing gaps decrease with increasing frequency. The reduction in pairing with frequency is due to the rotation and the gradual alignment of the paired nucleons. The pairing gaps of the ground-state band are larger than that of the two-particle bands. The reduction in pairing for the high-$K$ bands is due to the Pauli blocking of the orbitals near the Fermi surface.   

To examine the rotational frequency $\omega$ and seniority $\nu$ (number of the unpaired particles) dependences of the pairing gap quantitatively, the relative pairing gap reduction factors are defined as,
\begin{eqnarray}
\nonumber
R_{\tau}(\omega)&=&\frac{\tilde{\Delta}_{\tau}(\omega)-\tilde{\Delta}_{\tau}(\omega={0})}{\tilde{\Delta}_{\tau}(\omega={0})}, \\
R_{\tau}(\nu)&=&\frac{\tilde{\Delta}_{\tau}(\nu)-\tilde{\Delta}_{\tau}(\nu={0})}{\tilde{\Delta}_{\tau}(\nu={0})}, \ \ \ \ \ \ \ \ \ \ \ \ \tau=p\ \mathrm{or}\ n
\label{eq:PairingReduction}
\end{eqnarray} 
In the following studies, the seniority dependence of pairing gap $R_{\tau}(\nu)$ is calculated at the bandhead $\hbar\omega=0$, and the $\tilde{\Delta}_\tau (\nu = 0)$ is adopted as the $\tilde{\Delta}$ of GSB.     
\begin{eqnarray}
\nonumber
\mathrm{GSB}:&&\ \ \ \  
R_{p}(\omega=0.3 \mathrm{MeV}/\hbar)\approx18.1\%,\ \ \ \ \ \\\nonumber
\pi^{2}3^{+}:&&\ \ \ \  
R_{p}(\omega=0.3 \mathrm{MeV}/\hbar)\approx5.7\%,\ \ \ \ \ \ R_{p}(\nu=2)\approx4.5\%\\\nonumber
\pi^{2}8^{-}:&&\ \ \ \ 
R_{p}(\omega=0.3 \mathrm{MeV}/\hbar)\approx5.4\%,\ \ \ \ \ \ R_{p}(\nu=2)\approx4.4\%\\\nonumber
\label{eq:PairingReductionResult}
\\
\mathrm{GSB}:&&\ \ \ \ \nonumber
R_{n}(\omega=0.3 \mathrm{MeV}/\hbar)\approx22.3\%,\ \ \ \ \ \\\nonumber
\nu^{2}8^{-}_{1}:&&\ \ \ \  
R_{n}(\omega=0.3 \mathrm{MeV}/\hbar)\approx8.0\%,\ \ \ \ \ \ R_{n}(\nu=2)\approx4.2\%\\\nonumber
\nu^{2}10^{+}:&&\ \ \ \  
R_{n}(\omega=0.3 \mathrm{MeV}/\hbar)\approx8.0\%,\ \ \ \ \ \ R_{n}(\nu=2)\approx4.8\%.
\end{eqnarray}

The different behaviors of the observed GSB and high-$K$ bands in $^{254}$No can be explained. At the bandhead $\hbar\omega=0$, the seniority dependence of relative pairing gap reduction is about $\sim4.5\%$, which is due to the Pauli blocking of the unpaired nucleons occupying single-particle orbitals near the Fermi surface. This contributes to the $\sim25\%$ increases of $J^{(1)}$ for the high-$K$ (seniority $\nu=2$) bands compared with the ground-state  (seniority $\nu=0$) band. The frequency dependences of the relative pairing gap reduction at $\hbar\omega=0.3$ MeV are about $20\%$ for the GSB, and are about $5\%$ $(8\%)$ for two-proton (neutron) high-$K$ bands. Therefore, $J^{(1)}$ of two-particle $K^{\pi}=3^+,8^-$ and $10^+$ bands displays flat behavior while the GSB increases smoothly with frequency.  

\section{Summary}
The multi-particle states and rotational properties of two-particle  $K^{\pi}=3^+,8^-$ and $10^+$ bands in $^{254}$No have been investigated by the cranked shell model with pairing correlations treated by a particle-number conserving method. The experimental excitation energies and moments of inertia for the multi-particle state are reproduced well by the calculation. The calculated Nilsson levels with high-order deformation $\varepsilon_{6}$ show enlarged proton $Z=100$ and neutron $N=152$ deformed shell gaps. Better reproduction of the experimental data are achievied based on such single-particle levels structure. There is a signature splitting of the Nilsson proton orbital $\pi\frac{1}{2}^-[521]$. Only the state with configuration $\pi\frac{7}{2}^{-}[514](\alpha=+1/2)\otimes\pi\frac{1}{2}^{-}[521](\alpha=-1/2)$ can reproduce the experimental rotational behavior. The $J^{(1)}$ in two-particle state bands is larger than the ground-state band by about $25$\%. A detailed investigation into pairing shows that the rise of $J^{(1)}$ in two-particle state bands is attributed to the pairing reduction due to the Pauli blocking effects.  

\section*{Acknowledgments}
One of the authors, X.-T. He, is grateful to Prof. P. Walker for his very useful comments and for carefully reading the manuscript. This work is supported by the National Natural Science Foundation of China (Grant Nos. 11775112, 11535004 and 11761161001) and the Priority Academic Program Development of Jiangsu Higher Education Institutions.

\section*{References}
\bibliographystyle{apsrev4-1}
\bibliography{../../../../References/ReferencesXT}

\end{document}